\documentclass[aps,showpacs,nofootinbib,superscriptaddress]{revtex4}
\usepackage{graphicx}
\usepackage{dcolumn}

\def\slashchar#1{\setbox0=\hbox{$#1$}
   \dimen0=\wd0 \setbox1=\hbox{/} \dimen1=\wd1
   \ifdim\dimen0>\dimen1 \rlap{\hbox to \dimen0{\hfil/\hfil}} #1
   \else  \rlap{\hbox to \dimen1{\hfil$#1$\hfil}} / \fi}
\def\ereco{E_{\rm rec}}

\begin{document}
\title{Neutrino Energy Reconstruction and the Shape of the
  CCQE-like Total Cross Section}

\author{J. Nieves}
\affiliation{Instituto de F\'\i sica Corpuscular (IFIC), Centro Mixto
Universidad de Valencia-CSIC, Institutos de Investigaci\'on de
Paterna, E-46071 Valencia, Spain}
\author{F. S\'anchez}
\affiliation{Institut de Fisica d’Altes Energies (IFAE), Bellaterra
  Barcelona, Spain}
\author{I. \surname{Ruiz Simo}}
\affiliation{Departamento de F\'isica At\'omica, Molecular y Nuclear,  
Universidad de Granada, E-18071 Granada, Spain}
\author{M. J.  \surname{Vicente Vacas}}
\affiliation{Departamento de F\'\i sica Te\'orica and IFIC, Centro Mixto
Universidad de Valencia-CSIC, Institutos de Investigaci\'on de
Paterna, E-46071 Valencia, Spain}

\today

\begin{abstract}
 We show that because of the multinucleon mechanism effects, the
 algorithm used to reconstruct the neutrino energy is not adequate
 when dealing with quasielastic-like events, and a distortion of the
 total flux unfolded cross section shape is produced. This amounts to
 a redistribution of strength from high to low energies, which gives
 rise to a sizable excess (deficit) of low (high) energy
 neutrinos. This distortion of the shape leads to a good description
 of the MiniBooNE unfolded CCQE-like cross sections published in
 Ref.~\cite{AguilarArevalo:2010zc}. However, these changes in the
 shape are artifacts of the unfolding process that ignores
 multinucleon mechanisms.
\end{abstract}

\pacs{25.30.Pt,13.15.+g, 24.10.Cn,21.60.Jz}

\maketitle

\section{Introduction}

\begin{figure}[h]
\includegraphics[width=0.85\textwidth]{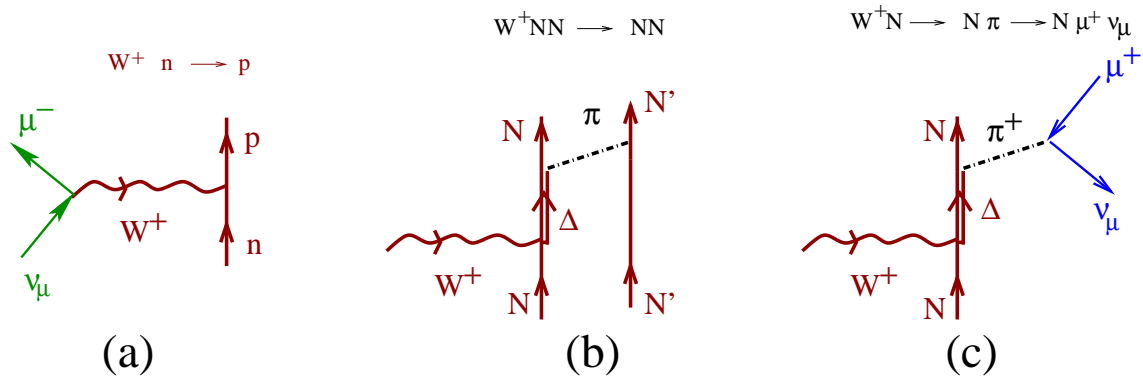}
\caption{\label{fig:expl}
Mechanisms for $W$ absorption inside of a nucleus.}
\end{figure}

In most theoretical works  the name Quasielastic (QE) scattering is used for processes
where the gauge boson $W$ is absorbed by just one nucleon, which
together with a lepton is emitted (see Fig.~\ref{fig:expl}(a)). However,
in the MiniBooNE measurement of Ref.~\cite{AguilarArevalo:2010zc}, QE
is related to processes in which only a muon is detected in the final
state. Though this definition could make sense because ejected
nucleons are not detected in that experiment, it includes
multinucleon processes (see Fig.~\ref{fig:expl}(b))\footnote{Note that
  the intermediate pion in this figure is virtual and it is part of
  the $\Delta N \to NN$ interaction inside of the nucleus. Indeed, one
  should consider a full interaction model for the in medium
  baryon--baryon interaction. Thus, for instance, the model of Ref.~\cite{Nieves:2011pp}
  contains, besides pion exchange, $\rho-$exchange, and short and long
  range (RPA) correlations.}  and others like pion production followed
by absorption. However, it discards pions coming off the nucleus,
since they will give rise to additional leptons after their decay (see
Fig.~\ref{fig:expl}(c)). The MiniBooNE analysis of the data corrects
(through a Monte Carlo estimate) for some of these events, where in the neutrino
interaction a real pion is produced, but it escapes detection because
it is reabsorbed in the nucleus, leading to multinucleon emission.

As firstly pointed out by M. Martini et
al.~\cite{Martini:2009uj,Martini:2010ex}, and corroborated by our
group ~\cite{Nieves:2011pp,Nieves:2011yp}, the data of
Ref.~\cite{AguilarArevalo:2010zc} correspond to the sum of the QE
(absorption by just one nucleon), and the multinucleon
contributions. For this reason, we will use the name QE-like to quote
the MiniBooNE data of Ref.~\cite{AguilarArevalo:2010zc}. Also for
simplicity, we will often refer to the multinucleon mechanism
contributions, though they include effects beyond gauge boson
absorption by a nucleon pair, as 2p2h (two particle-hole)
effects. The 2p2h contributions allows to
describe~\cite{Nieves:2011yp,Martini:2011wp} the CCQE-like flux
averaged double differential cross section $d\sigma/dE_\mu
d\cos\theta_\mu$ measured by MiniBooNE with values of $M_A$ (nucleon
axial mass) around $1.03\pm 0.02$ GeV that is usually quoted as the
world average~\cite{Bernard:2001rs,Lyubushkin:2008pe}. This is
re-assuring from the theoretical point of view and more satisfactory
than the situation envisaged by some other works that described these
CCQE-like data in terms of a larger value of $M_A$ of around 1.3--1.4
GeV~\cite{AguilarArevalo:2010zc,Benhar:2010nx,Juszczak:2010ve,
  Butkevich:2010cr}.  

  For the QE cross-sections, the predictions of the model that we employed in
  \cite{Nieves:2011yp} agree quite well with those obtained/used in
  Refs.~\cite{Martini:2009uj,Martini:2010ex, Martini:2011wp}, and 
  both groups also agree on the relevant role played by the
  2p2h mechanisms to describe the MiniBooNE data. We, however, differ
  considerably in the size (about a factor of two) of the multinucleon
  effects~\cite{Nieves}. Thus, although Martini et al., predictions
  look consistent with MiniBooNE data, however our predictions, when
  the 2p2h contribution is included, would favor a global
  normalization scale of about 0.9 (see \cite{Nieves:2011yp}). This
  would be consistent with the MiniBooNE estimate of a total
  normalization error of 10.7\%. In view of the disagreement, we
  should emphasized here that our evaluation in
  ~\cite{Nieves:2011pp,Nieves:2011yp}, of these pionless multinucleon
  emission contributions to the cross section is fully microscopical
  and it contains terms, which were either not considered or only
  approximately taken into account in \cite{Martini:2009uj,
    Martini:2010ex, Martini:2011wp}. Indeed, the results of these
  latter works rely on some computation of the 2p2h mechanisms for the
  $(e,e')$ inclusive reaction (\cite{Alberico:1983zg}), whose results
  are used for neutrino induced processes. Thus, it is clear that
  these latter calculations do not contain any information on axial or
  axial-vector contributions.

We would also like to point out that the simple phenomenological
approach adopted in \cite{Lalakulich:2012ac} to account for the 2p2h
effects also reinforces the picture that emerges from the works of
Refs.~\cite{Nieves:2011yp,Martini:2011wp}. Yet, a partial
microscopical calculation of the 2p2h contributions to the CCQE cross
section has been also presented in Refs.~\cite{Amaro:2010sd} and
\cite{Amaro:2011aa}, for neutrino and antineutrino induced reactions,
respectively. In these works, the contribution of the vector meson
exchange currents in the 2p2h sector is added to the QE neutrino or
antineutrino cross section predictions deduced from a phenomenological
model (SuSA)~\cite{Amaro:2004bs} based on the super-scaling behavior
of electron scattering data.  In \cite{Amaro:2011qb}, and for the
neutrino case, the SuSA+2p2h results were also compared with those
obtained from a relativistic mean field approach. Although, all these
schemes do not account for the axial part of the 2p2h effects yet, the
preliminary results also corroborate that 2p2h meson exchange currents
play an important role in both CCQE neutrino and antineutrino
scattering, and that they may help to resolve the controversy on the
nucleon axial mass raised by the recent MiniBooNE data. This is not
surprising, since these two-body currents, that arise from microscopic
relativistic modeling performed for inclusive electron scattering
reactions, are known to result in a significant increase in
the vector–-vector transverse response function in  QE
electron scattering
data~\cite{DePace:2003xu,Amaro:2010iu,Amaro:2009dd}.

The study and comparison in detail of the different models used to
describe 2p2h effects, though of great interest, is left for future
research. We aim here at determining the possible influence of the
2p2h excitations on the process needed to extract neutrino energy
unfolded cross sections from the measured flux-average data.

Indeed, there still exists another feature of neutrino physics which deserves
attention and that has motivated this work. Neutrino beams are not
monochromatic. For QE-like events, only the charged lepton is
observed and the only measurable quantities are then its
direction\footnote{From now on, we will always identify the charged
  lepton with a muon.}
(scattering angle $\theta_\mu$ with respect to the neutrino beam
direction) and its energy $E_\mu$. The energy of the neutrino that has
originated the event is unknown. Then, it is common to define a 
reconstructed neutrino energy $\ereco$ as,
\begin{equation}
\ereco = \frac{M
  E_\mu-m_\mu^2/2}{M-E_\mu+|\vec{p}_\mu|\cos\theta_\mu}\label{eq:defereco}
\end{equation}
which will correspond to the energy of a neutrino that emits a muon,
of energy $E_\mu$ and three-momentum $\vec{p}_\mu$, and a gauge boson
$W$ that is being absorbed by a nucleon of mass $M$ at rest. Namely,
the usual reconstruction procedure assumes that we are dealing with a
genuine quasielastic event on a nucleon at rest, ie. $\ereco$ is
determined by the QE-peak condition $q^0=-q^2/2M$, where $q^\mu$ is
the $W$ four momentum.  Note that each event contributing to the flux
averaged double differential cross section $d\sigma/dE_\mu
d\cos\theta_\mu$ defines unambiguously a value of $\ereco$.  The
actual (``true'') energy, $E$, of the neutrino that has produced the
event will not be exactly $\ereco$.  Actually, for each $\ereco$,
there exists a distribution of true neutrino energies that could give
rise to events whose muon kinematics would lead to the given value of
$\ereco$. Several effects can influence this distribution. Firstly,
the Fermi motion which broadens the QE peak and the Pauli blocking
which cuts the low momentum response. These effects are well
known\footnote{It is also common to consider some corrections in the
  definition of $\ereco$ in Eq.~(\ref{eq:defereco}) to account for the
  binding energy of the target nucleon in the nucleus, but these
  corrections turn out to be irrelevant for our discussion.}, usually
are under control and lead to very minor changes in the process of
expressing observables as a function of the true neutrino energy.  This
is because genuine QE events produce true energy distributions quite
narrow and strongly peaked around the expected $\ereco$
values~\cite{Martini:2012fa}.  However multinucleon mechanisms,
relevant for QE-like processes, can indeed distort the expected
(QE-based) $(\ereco, E)$ distributions, since they produce
distributions quite flat and that do not peak around $\ereco$
\cite{Martini:2012fa}.  The effects of the inclusion of multinucleon
processes on the energy reconstruction have been investigated in
Ref.~\cite{Martini:2012fa}, within their 2p2h model and also estimated
in Ref.~\cite{Mosel:2012hr}, using some simplified model for the
multinucleon mechanisms.

We will show in this work that 2p2h effects sizably distort the shape
of the total CCQE-like flux unfolded cross section, as a function of
the neutrino energy. Indeed, we will see that these multinucleon
mechanisms produce a redistribution of strength from high energy to
low energies, which gives rise to a sizable enhancement of the number
of events attributed to low energy neutrinos leading to a good
description of the unfolded cross section given in
\cite{AguilarArevalo:2010zc}.  However, we will show that these
changes in the shape are artifacts of the unfolding process that
ignores multinucleon mechanisms.

\section{Excess of low energy neutrinos in the 
MiniBooNE CCQE-like flux unfolded cross section  data}

The QE+multinucleon mechanism model of
Ref.~\cite{Nieves:2011yp}, with $M_A=1.049$ GeV, provides an excellent
description of the MiniBooNE neutrino flux folded CCQE-like
$d\sigma/dT_\mu d\cos\theta_\mu$ differential cross sections given in
~\cite{AguilarArevalo:2010zc}, even when no
parameters have been fitted to data, beyond a global scale,
$\lambda\sim 0.9$. This scale $\lambda$ is consistent with
the global normalization uncertainty of around 10\% acknowledged in
~\cite{AguilarArevalo:2010zc}.

The QE contribution used in ~\cite{Nieves:2011yp} was derived in
Ref.~\cite{Nieves:2004wx}, and it incorporates several nuclear effects. The
main one is the medium polarization (RPA), including $\Delta$-hole
degrees of freedom and explicit $\pi$ and $\rho$ meson exchanges in
the vector-isovector channel of the effective nucleon-nucleon
interaction. The model for multinucleon mechanisms has been fully
discussed in Ref.~\cite{Nieves:2011pp} and it is based on a model for
neutrino pion production derived in
Refs.~\cite{Hernandez:2007qq,Hernandez:2010bx}. The whole model
constitutes a natural extension of previous studies of photon,
electron, and pion interactions with
nuclei~\cite{Gil:1997bm,Carrasco:1989vq,Nieves:1991ye,Nieves:1993ev}.

\begin{figure}
\makebox[0pt]{\includegraphics[width=0.55\textwidth]{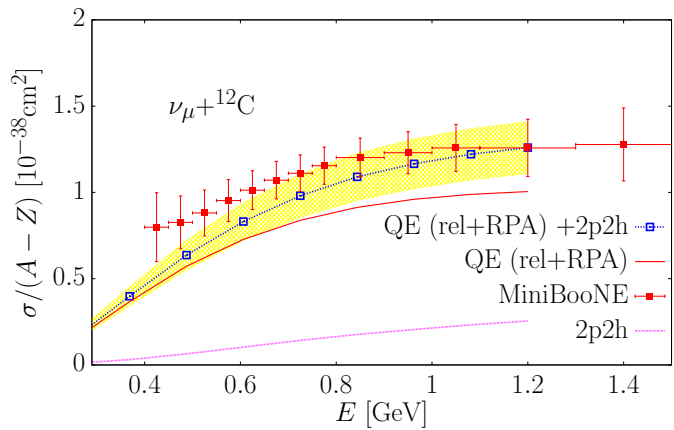}}
\caption{Flux-unfolded MiniBooNE $\nu_\mu$ CCQE-like cross section per
  neutron as a function of neutrino energy (data points) from
  Ref.~\cite{AguilarArevalo:2010zc}, together with the predictions
  derived from the model used in Ref.~\cite{Nieves:2011yp}. The yellow
  band accounts for theoretical uncertainties, as discussed in
  Ref.~\cite{Nieves:2011pp}, while the QE contribution includes
  relativistic effects, and some nuclear corrections, among other
  those due to long range RPA correlations. }
\label{fig:ccqe-like-th}
\end{figure}

The prediction of the model, used in Ref.~\cite{Nieves:2011yp}, for
the total flux unfolded neutrino CCQE-like cross section is depicted
in Fig.~\ref{fig:ccqe-like-th}. Several remarks are in order here:
\begin{itemize}

\item The 2p2h contributions clearly improve the description of the
  data in Fig.~\ref{fig:ccqe-like-th}, which are totally missed by the
  QE prediction. Though the model provides a reasonable description,
  we observe a sizable excess of low energy neutrinos in the data,
  that is not even covered by the theoretical error band.

\item As discussed above, the flux folded double differential cross-section
  data is well described in Ref.~\cite{Nieves:2011yp}, except for 
  the global scale, $\lambda\sim 0.9$, which needs to be introduced
  there. It is to say, predicted cross sections in
  Ref.~\cite{Nieves:2011yp} are globally around
  10\% smaller than the measured ones. This disagreement could be  
  due to a theoretical under-estimation of the absolute number of 
  neutrinos in the MiniBooNE flux. Thus, we should expect our
  predictions for the total unfolded cross section to under-estimate
  the data points by about 10\%, as well. However, we do not see this
  in Fig.~\ref{fig:ccqe-like-th}. There is a problem in the
  neutrino-energy shape, as pointed out above, which would not be
  improved by increasing the size our predictions by a global factor.

  In any case, given that the actually measured quantity it is the
  double differential cross-section and that observable is well
  described by our model, any difference on the unfolded cross section
  must come from the unfolding procedure.

\item Finally, we should mention that the QE theoretical results for
  the cross section shown here (model from ~\cite{Nieves:2011yp})
  slightly differ from those in Fig.~18 (left) of our previous work of
  Ref.~\cite{Nieves:2011pp}. The main difference is the inclusion of
  relativistic corrections. We discuss this point in some detail in
  Appendix~\ref{sec:app}, since we do not want to deviate here the
  attention from the main point of this work: because of the 2p2h
  effects, the algorithm used to reconstruct the neutrino energy is
  not adequate when dealing with QE-like events, and that it produces
  a distortion of the total CCQE-like flux unfolded cross section
  shape. 

Nevertheless, we should mention here that in Ref.~\cite{Nieves:2011pp} and
to account for Final State Interactions (FSI), we used the
non-relativistic QE model of Ref.~\cite{Nieves:2004wx}, while in
Fig.~\ref{fig:ccqe-like-th}, the results depicted are those obtained
from the relativistic model of ~\cite{Nieves:2004wx} for the QE
process, without the inclusion of FSI effects\footnote{In
  Ref.~\cite{Nieves:2011pp}, FSI effects are being treated within the
  non-relativistic scheme derived in
  Ref.~\cite{FernandezdeCordoba:1991wf}. A non-relativistic treatment
  is unsuitable for the large momenta transferred that are reached in
  the MiniBooNE neutrino flux folded $d\sigma/dT_\mu d\cos\theta_\mu$
  differential cross sections, but it is more appropriated for the
  total unfolded cross section as long as the neutrino energy is
  sufficiently small. A final consideration, in
  Ref.~\cite{Benhar:2010nx}, it was found that the main effect of FSI
  is a shift of $\sim 10$ MeV of the QE peak for neutrino energies
  closer to the MiniBooNE neutrino flux mean energy, $\langle E
  \rangle \sim 800 $ MeV and that it has little impact on the
  integrated cross section (we will illustrate, within the model of
  Ref.~\cite{Nieves:2004wx}, this latter affirmation also in the
  Appendix~\ref{sec:app}).  However, we cannot discard the possibility
  that these effects could be more important in the angle and energy
  distributions for low energy neutrinos. Moreover, it has been also
  pointed out that some relativistic approaches to account for FSI
  lead to larger variations of the total cross
  section~\cite{Meucci:2011vd}. }. This improvement in the model to
account for the relativistic effects is the reason for the differences
mentioned above. As a final remark, we stress that the results of
Ref.~\cite{Nieves:2011yp} for the double differential cross section
and those displayed in Fig.~\ref{fig:ccqe-like-th} have been
calculated with the same model.

\end{itemize}
We see in Fig.~\ref{fig:ccqe-like-th} that the proportion of
multinucleon events contributing to the QE-like signal is quite large
in the whole energy range relevant in the MiniBooNE experiment.  This
questions the validity of the algorithm used to reconstruct the 
neutrino energy in Eq.~(\ref{eq:defereco}). In the next section, we will
explore in detail this problem and we will  find out the shape
distortion effects induced by the use of Eq.~(\ref{eq:defereco}) in the
MiniBooNE data.
\section{The distribution of reconstructed energies in CCQE-like experiments}

Let $P(\ereco^0|E)$ be the conditional probability density
of measuring an event with reconstructed energy (combination of muon
energy and scattering angle given in Eq.~(\ref{eq:defereco}))
comprised in the interval $[\ereco^0,\ereco^0+d\ereco]$ and induced by
    the interaction with the nuclear target of a neutrino of energy
    $E$ (it is to say, conditional probability density of obtaining $\ereco$
    ``given'' $E$). This probability density can be computed theoretically as:
\begin{equation}
P(\ereco^0| E) = \frac1{\sigma(E)}
\frac{d\sigma}{d\ereco}(E;\ereco=\ereco^0)  \label{eq:pcon}
\end{equation}
where $\sigma(E)$
is the integrated CCQE-like cross section for neutrinos of energy $E$,
and the distribution $d\sigma/d\ereco$ is obtained from the double
differential cross section, with respect the energy and scattering angle of the
outgoing muon, as
\begin{equation}
\frac{d\sigma}{d\ereco}(E;\ereco^0) = \int_{m_\mu}^E dE_\mu
\frac{d^2\sigma }{d\ereco dE_\mu}(E;\ereco^0)= \int_{m_\mu}^E dE_\mu
\left|\frac{\partial(\cos\theta_\mu)}{\partial\ereco}\right|
\frac{d^2\sigma }{d(\cos\theta_\mu) dE_\mu}(E;\ereco^0) \label{eq:defsigreco}
\end{equation}
for a fix value of the reconstructed energy $\ereco=\ereco^0$ and
``true'' neutrino energy $E$. Eq.~(\ref{eq:defereco}) can be used to
express $\cos\theta_\mu$ in terms of $E_\mu$ and $\ereco$. Besides,
the Jacobian can be trivially computed also from
Eq.~(\ref{eq:defereco}) and it reads
\begin{equation}
\frac{\partial(\cos\theta_\mu)}{\partial\ereco} = - \frac{M
  E_\mu-m_\mu^2/2}{\ereco^2 |\vec{p}_\mu|}
\end{equation}
We would like to stress that $d\sigma/d\ereco(E;\ereco^0)$ is an observable, but
this distribution is not accessible to experiments where only the
kinematics of the outgoing muon is measured.

On the other hand, let $P_{\rm rec}(\ereco)$ be the 
probability density of measuring an event with reconstructed energy
$\ereco$, 
\begin{equation}
P_{\rm rec}(\ereco) = \int P(\ereco | E) P_{\rm true} (E) dE
\end{equation}
where
\begin{equation}
P_{\rm true}(E) = \frac{1}{ \langle
\sigma \rangle} \Phi(E) \sigma(E), \qquad  
\langle\sigma \rangle= \int\Phi(E') \sigma(E') dE' \label{eq:ptrue}
\end{equation}
is the density probability of having an event due to the interaction
of a neutrino, with energy  between $E$ and  $E+dE$, with the nuclear
target. $\Phi$ is the neutrino flux normalized to one, and
$\langle\sigma \rangle$ is the total flux averaged cross section. It trivially
follows, 
\begin{equation}
P_{\rm rec}(\ereco) = \frac{1}{ \langle \sigma \rangle} 
\int  \frac{d\sigma}{d\ereco}(E;\ereco) \Phi(E) dE \label{eq:prec}
\end{equation}
a magnitude that can be measured in a CCQE-like experiment.

\section{The distribution of true neutrino energies $P_{\rm true}(E)$ from CCQE-like experiments}

We use Bayes's theorem to estimate $P_{\rm true}(E)$ from the
measured density probability $P_{\rm rec}(\ereco)$. To that end, let
us introduce $P(E|\ereco)$ that is, given an event of
reconstructed energy $\ereco$, the conditional density probability of being
produced by a neutrino of energy $E$. It follows, 
\begin{equation}
P_{\rm true}(E) = \int d\ereco P_{\rm rec}(\ereco)P(E|\ereco)
\end{equation}
Now, since Bayes's theorem reads
\begin{equation}
P(E| \ereco ) = \frac{P(\ereco| E) P_{\rm true}(E)
}{P_{\rm rec}(\ereco) }
\label{eq:bayes} 
\end{equation}
we deduce 
\begin{equation}
P(E| \ereco ) = \frac{\Phi(E)d\sigma/d\ereco(E;\ereco)}{\int
dE''\Phi(E'')d\sigma/d\ereco(E'';\ereco)}\label{eq:pdec}
\end{equation}
from Eqs.~(\ref{eq:pcon}), (\ref{eq:ptrue}) and (\ref{eq:prec}). The
recent work of M. Martini et al. \cite{Martini:2012fa} pays an special
attention to this distribution that, as  we observe,  depends on the
neutrino flux. The
above equation implies
\begin{equation}
P_{\rm true}(E) = \int d\ereco P_{\rm rec}(\ereco)\frac{\Phi(E)d\sigma/d\ereco(E;\ereco)}{\int
dE''\Phi(E'')d\sigma/d\ereco(E'';\ereco)}
\end{equation}
Finally and attending to the existing relation between $P_{\rm true}(E)$
and $\sigma(E)$ in Eq.~(\ref{eq:ptrue}), we could write
\begin{equation}
\sigma(E) = \int d\ereco \Big [\langle \sigma \rangle P_{\rm
    rec}(\ereco)\Big] \times \left[ \frac{d\sigma/d\ereco(E;\ereco)}{\int
dE''\Phi(E'')d\sigma/d\ereco(E'';\ereco)}\right] \label{eq:defsigma}
\end{equation}
A consistency check is obtained if we substitute
Eq.~(\ref{eq:prec}) in Eq.~(\ref{eq:defsigma}) which leads to
\begin{equation}
\sigma(E) = \int d\ereco \frac{d\sigma}{d\ereco}(E;\ereco)
\end{equation}
that it is trivially satisfied thanks to the definition of
$d\sigma/d\ereco(E;\ereco)$ in Eq.~(\ref{eq:defsigreco}).

Eq.~(\ref{eq:defsigma}) might be used to
estimate the integrated flux unfolded cross section from data. CCQE-like experiments
measure the quantities that appear in the first  bracket of this
equation, namely, $\langle \sigma \rangle \times P_{\rm
    rec}(\ereco)$. We have already discussed about $P_{\rm
    rec}(\ereco)$, while the total flux averaged cross section
$\langle \sigma \rangle$  is determined by the ratio of the total number of
events ($N_{event}$) over the number of incident neutrinos per unit
of area ($N_{inc}$). $N_{event}$ is directly measured while for
$N_{inc}$ there exist, in principle, accurate theoretical
predictions. Note that the flux $\Phi(E)$, normalized to one,
gives only the shape of the neutrino flux, but it is independent of the total number 
of incident neutrinos. 

Thus, if one had a theoretical model for the second bracket $\left[
  d\sigma/d\ereco(E;\ereco)/\int
  dE''\Phi(E'')d\sigma/d\ereco(E'';\ereco)\right]$ in
Eq.~(\ref{eq:defsigma}) or equivalently for $P(E| \ereco )$, one could
extract the flux unfolded cross section $\sigma(E)$ after folding it
with the measured data ($\langle \sigma \rangle \times P_{\rm
  rec}(\ereco)$).  This method reproduces~\cite{miniboone} the data unfolding used by
the MiniBooNE collaboration in Ref.~\cite{AguilarArevalo:2010zc} and
described in Ref.~\cite{Agostini:1995}. The iterative unfolding method
in MiniBooNE is needed due to the statistical fluctuations in the data
and the lack of "a priori" knowledge of the $P_{true}(E)$ probability
in Eq.~(\ref{eq:bayes}), both are not relevant for theoretical
calculations. However, as a proof of the validity of this approach we
have checked that the iterative method in \cite{Agostini:1995} yields
to identical results.  As discussed in the introduction, the works of
Refs.~\cite{Nieves:2011pp,Martini:2009uj, Martini:2010ex,
  Nieves:2011yp,Martini:2011wp} show that the QE-like (and/or
differential) cross section is given by the sum
\begin{equation}
\sigma(E) = \sigma^{\rm QE}(E) + \sigma^{\rm 2p2h}(E)
\end{equation}
of the genuine QE and the multinucleon contributions. Up to now,
experimental analysis have completely neglected the latter (2p2h)
cross section,
while well established nuclear corrections, like RPA correlations,
have also been  ignored in the computation of the former one (QE). As a consequence, a high
value of $M_A> 1.3$ GeV is required in the MiniBooNE analysis
 to describe  the flux folded $d\sigma/dq^2$ or
$d\sigma/dT_\mu d\cos\theta_\mu$
 distributions~\cite{AguilarArevalo:2010zc}. But, if this approximate
 model is used in Eq.~(\ref{eq:defsigma}) to extract the unfolded
 cross section,
\begin{equation}
\sigma_{\rm appx}(E) = \int d\ereco \Big [\langle \sigma \rangle P_{\rm
    rec}(\ereco)\Big]_{\rm Exp} \times \left[ \frac{d\sigma/d\ereco(E;\ereco)}{\int
dE''\Phi(E'')d\sigma/d\ereco(E'';\ereco)}\right]_{{\rm QE~ no~
    RPA,}~M_A> 1.3~{\rm GeV}} \label{eq:poormodel}
\end{equation}
the resulting estimate $\sigma_{\rm appx}(E)$ might significantly differ
from the real QE-like neutrino-nucleus cross section
$\sigma(E)$. Actually, we will show that a redistribution of strength
from high energy to low neutrino energies is being produced. To
illustrate this, we  will focus on the total cross section data on
carbon published by the MiniBooNE in \cite{AguilarArevalo:2010zc}.
We take in Eq.~(\ref{eq:poormodel})
\begin{equation}
\Big [\langle \sigma \rangle P_{\rm
    rec}(\ereco)\Big]_{\rm Exp} \sim \int
\left(\frac{d\sigma}{d\ereco}(E';\ereco)\Big|_{{\rm QE+RPA,}}^{M_A=
  1.049~{\rm GeV}}+ \frac{d\sigma^{\rm
    2p2h}}{d\ereco}(E';\ereco)\right) \Phi(E') dE' \label{eq:thmodel}
\end{equation}
where we have used Eq.~(\ref{eq:prec}) with our best theoretical
model, since we should mimic the experiment. Indeed, this is the same
model we employed in Ref.~\cite{Nieves:2011yp} to successfully
describe the MiniBooNE CCQE-like flux averaged double differential
cross section $d\sigma/dE_\mu d\cos\theta_\mu$, up to a global
normalization scale $\lambda$ ($=0.89\pm 0.01$). This latter parameter
is the only one which is fitted to data ($\chi^2/dof=53/137$). There
are no free parameters in the description of nuclear effects, since
they were fixed in previous studies of photon, electron, and pion
interactions with nuclei
\cite{Gil:1997bm,Carrasco:1989vq,Nieves:1991ye,Nieves:1993ev,Oset:1981ih,Salcedo:1987md}. Besides,
form factors are determined in independent analysis of the
experimental data on nucleons. In particular, the model uses a value
for the nucleon axial mass of $M_A=1.049$ GeV, that agrees within
errors with the world average~\cite{Bernard:2001rs,Lyubushkin:2008pe}
value of $1.03\pm 0.02$ GeV. As already mentioned, the genuine QE
piece was computed in Ref.~\cite{Nieves:2011pp}, it includes
relativistic effects, and some other nuclear corrections in addition
to the RPA ones, explicitly included in the label of
Eq.~(\ref{eq:thmodel}), though it does not include FSI effects. The
multinucleon cross section is taken from \cite{Nieves:2011pp}. The QE
and 2p2h contributions to the integrated cross section $\sigma(E)$ in
carbon were displayed in Fig.~\ref{fig:ccqe-like-th}. For simplicity,
in what follows, we will label this QE model as ``QE (rel+RPA)'' as in
that figure.

On the other hand,
to compute the second factor in the right hand side of
Eq.~(\ref{eq:poormodel}) we need to mimic the input used in the
experimental analysis. To that end,  we use a simple Fermi gas model with
$M_A=1.32$ GeV that only accounts for the genuine QE contribution and 
that does not include RPA corrections. The value of
$M_A$ in this model was fitted to the flux-folded double
differential  cross section $d\sigma/dE_\mu  d\cos\theta_\mu$ in
Ref.~\cite{Nieves:2011yp}, leading to an excellent value of
$\chi^2/dof=35/137$. This model should be quite similar to the one
originally used in the MiniBooNE analysis. The main
difference being that we use a local rather than global Fermi gas
in the calculation. Gathering the different terms, we have
\begin{eqnarray}
\sigma_{\rm appx}(E) &\sim &\sigma_{\rm appx}^{\rm QE
  ~(rel+RPA)}(E)+\sigma_{\rm appx}^{\rm 2p2h } (E) \label{eq:poormodel2}\\
\nonumber\\
\sigma_{\rm appx}^{\rm QE
  ~(rel+RPA)}(E) &\sim& \int d\ereco \int
\overbrace{\frac{d\sigma^{\rm QE
  ~(rel+RPA)}}{d\ereco}(E';\ereco)}^{M_A=
  1.049~{\rm GeV}} \Phi(E') dE' \times \overbrace{\left[ \frac{d\sigma/d\ereco(E;\ereco)}{\int
dE''\Phi(E'')d\sigma/d\ereco(E'';\ereco)}\right]}^{{\rm QE~ no~
    RPA,}~M_A=1.32~{\rm GeV}} \label{eq:poormodel2a}\\
\nonumber \\
\nonumber \\
\sigma_{\rm appx}^{\rm 2p2h}(E) &\sim& \int d\ereco \int \frac{d\sigma^{\rm
    2p2h}}{d\ereco}(E';\ereco)\Phi(E') dE'\times \underbrace{\left[ \frac{d\sigma/d\ereco(E;\ereco)}{\int
dE''\Phi(E'')d\sigma/d\ereco(E'';\ereco)}\right]}_{{\rm QE~ no~
    RPA,}~M_A=1.32~{\rm GeV}} \label{eq:poormodel2b}
\end{eqnarray}
\begin{figure}
\includegraphics[width=0.55\textwidth]{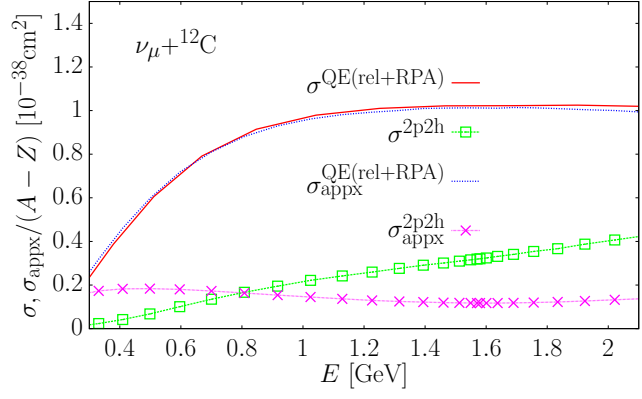}
\caption{Theoretical ($\sigma$) and approximate ($\sigma_{\rm appx}$,
  defined in Eqs.~(\ref{eq:poormodel2})--(\ref{eq:poormodel2b})) CCQE-like integrated cross sections in carbon as a
function of the neutrino energy.}
\label{fig:sigappx}
\end{figure}

\begin{figure}
\makebox[0pt]{\includegraphics[width=0.55\textwidth]{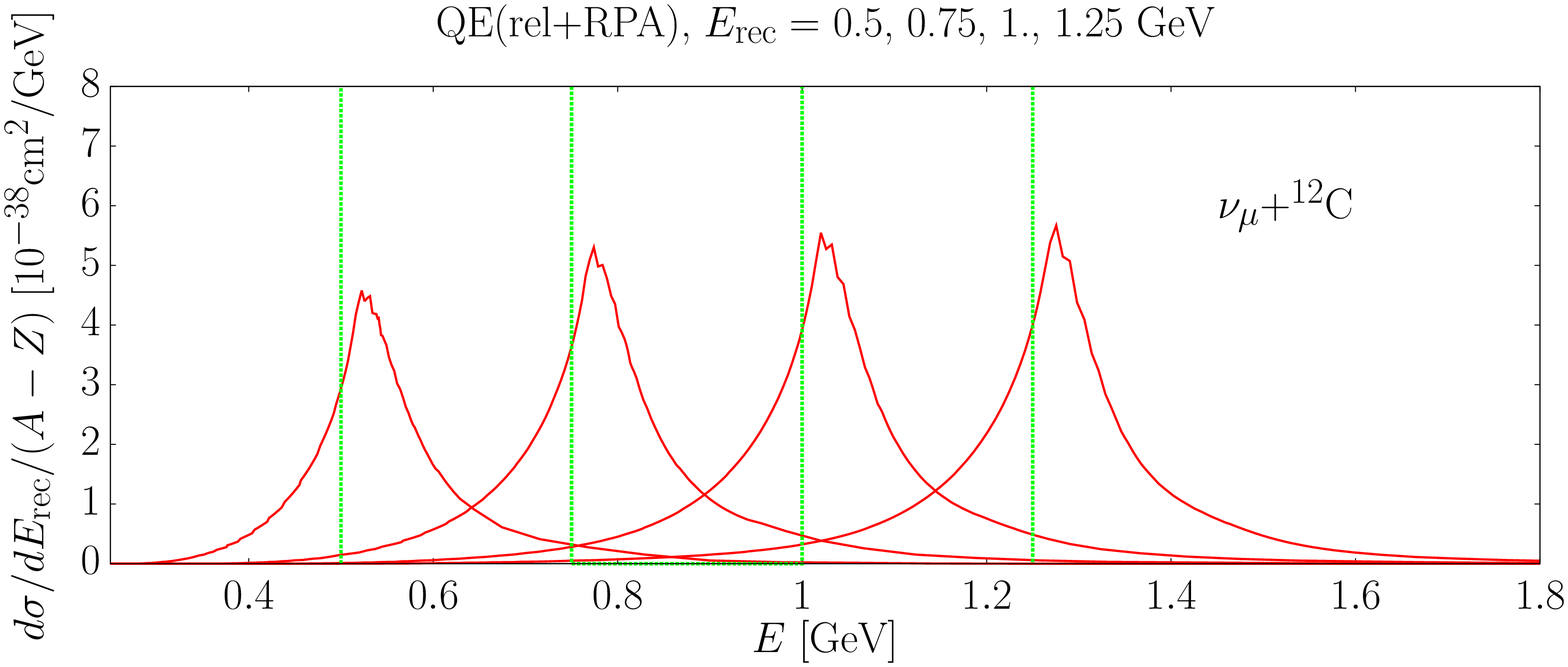}}\\
\vspace{0.5cm}
\makebox[0pt]{\includegraphics[width=0.55\textwidth]{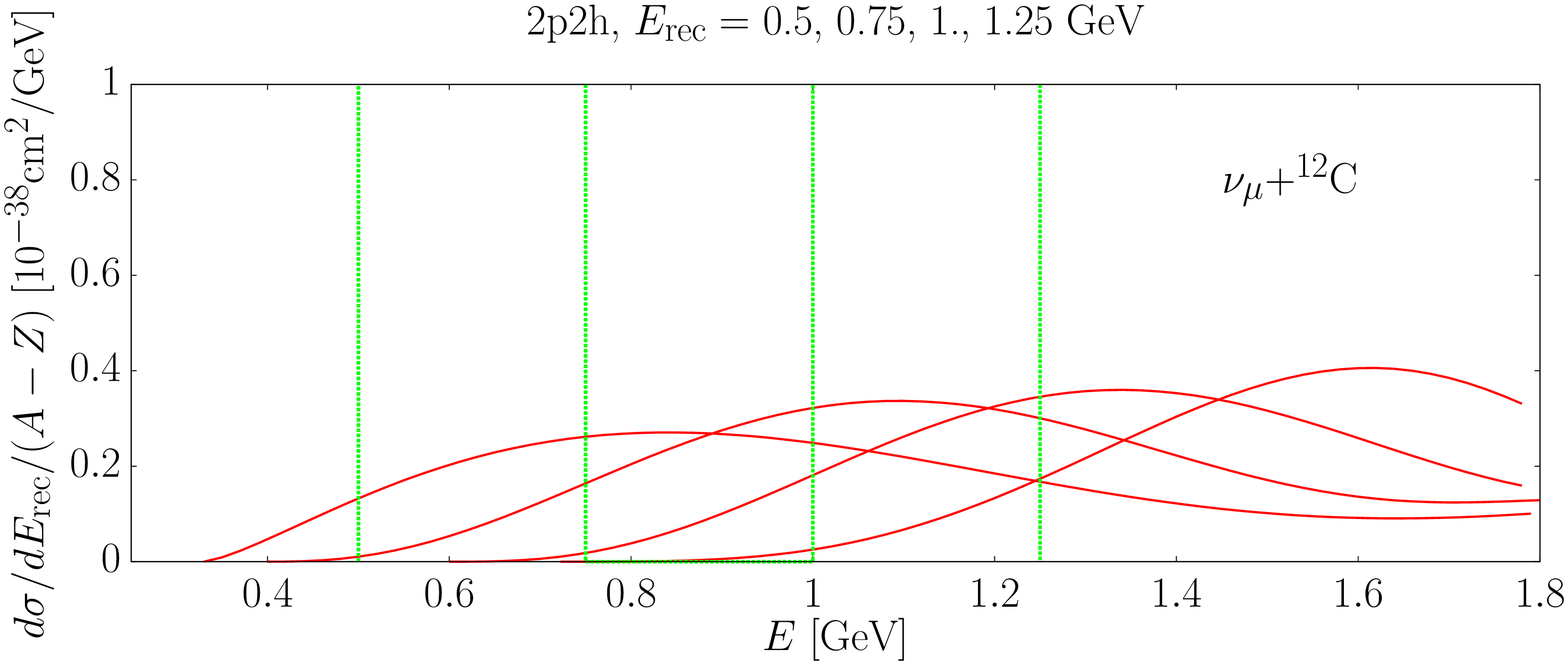}}
\caption{Differential cross section $d\sigma/d\ereco (E;\ereco)$ as a
  function of the ``true'' neutrino energy ($E$) and  four different
  values of $\ereco=0.5,0.75,1$ and 1.25 GeV, which are indicated by
  the vertical lines. QE contributions, from the (rel+RPA) model, are
  displayed in top panel, while 2p2h ones are shown in the bottom
  plot.}
\label{fig:dsigdereco}
\end{figure}

Results both for the QE  and the 2p2h contributions  to $\sigma_{\rm appx}$ are
shown in Fig.~\ref{fig:sigappx}. As can be
appreciated there, $\sigma_{\rm appx}(E)$ is an excellent approximation
to the real $\sigma (E)$ cross section in the case of the QE
contribution. The reason is that for genuine QE processes, 
the distribution $d\sigma^{\rm QE}/d\ereco(E;\ereco)$ is strongly peaked around
$E \approx \ereco$, as seen in the top panel\footnote{Actually the peak is
  shifted about 25 MeV up to higher energies since the bound energy of the target nucleon is not considered in Eq.(\ref{eq:defereco}). On the other hand, RPA correlations
  modify the size but do not affect significantly this peak structure.} of
Fig.~\ref{fig:dsigdereco}.\footnote{If for illustration purposes, we use
a Dirac's delta to approximate
%
$\frac{d\sigma^{\rm QE}}{d\ereco}(E;\ereco) \approx \sigma^{\rm QE}(E)\times
\delta(E-\ereco) \label{eq:approx}$
%
then, we will have
\begin{equation}
\overbrace{\left[ \frac{d\sigma/d\ereco(E;\ereco)}{\int
dE''\Phi(E'')d\sigma/d\ereco(E'';\ereco)}\right]}^{{\rm QE~ no~
    RPA,}~M_A=1.32~{\rm GeV}} \approx \frac{\delta(E-\ereco)}{\Phi(\ereco)}
\end{equation}
and therefore, independently of the nuclear model for QE, within this limit
\begin{equation}
\sigma_{\rm appx}^{\rm QE
  ~(rel+RPA)}(E) \approx \int d\ereco
\frac{\delta(E-\ereco)}{\Phi(\ereco)}  \int
\overbrace{\sigma^{\rm QE (rel+RPA)}(E')}^{M_A=
  1.049~{\rm GeV}} \delta(E'-\ereco)\Phi(E') dE' = \sigma^{\rm QE
  ~(rel+RPA)}(E)
\end{equation}
}

Though the actual $d\sigma^{\rm QE}/d\ereco$ distributions have some
widths, these differential cross sections are sufficiently peaked to
render the width effects on the ratio $\left(\sigma_{\rm appx}^{\rm QE
  ~(rel+RPA)}/\sigma^{\rm QE  ~(rel+RPA)}\right)$ quantitatively
irrelevant, as the results of the  Fig.~\ref{fig:sigappx}
indicate. Thus, we could conclude that when dealing only with genuine QE
events the procedure outlined in Eq.~(\ref{eq:poormodel}) to obtain the
flux unfolded cross section is quite accurate. This is
despite of the fact that RPA correlations and other nuclear effects were not considered
in the ansatz for  $P(E| \ereco )$ (second
bracket in Eq.~(\ref{eq:defsigma})). Note however, that all nuclear effects
are included in the first factor   $\Big [\langle \sigma \rangle P_{\rm
    rec}(\ereco)\Big]_{\rm Exp}$ in Eq.~(\ref{eq:poormodel}).

However, the situation is drastically different for the
2p2h contribution case, as one can also observe  in 
Fig.~\ref{fig:sigappx}. Indeed, it turns out that $\sigma_{\rm
  appx}^{\rm 2p2h}(E)$ is a poor estimate of the actual multinucleon
mechanism contribution $\sigma^{\rm 2p2h}(E)$. As before, if we
approximate
\begin{equation}
\sigma_{\rm appx}^{\rm 2p2h}(E) \approx \int d\ereco
\frac{\delta(E-\ereco)}{\Phi(\ereco)}  \int \frac{d\sigma^{\rm
    2p2h}}{d\ereco}(E';\ereco)
\Phi(E') dE' = \frac{1}{\Phi(E)} \int \frac{d\sigma^{\rm
    2p2h}}{d\ereco}(E';\ereco=E)
\Phi(E') dE'
\end{equation}
\begin{figure}
\includegraphics[width=0.55\textwidth]{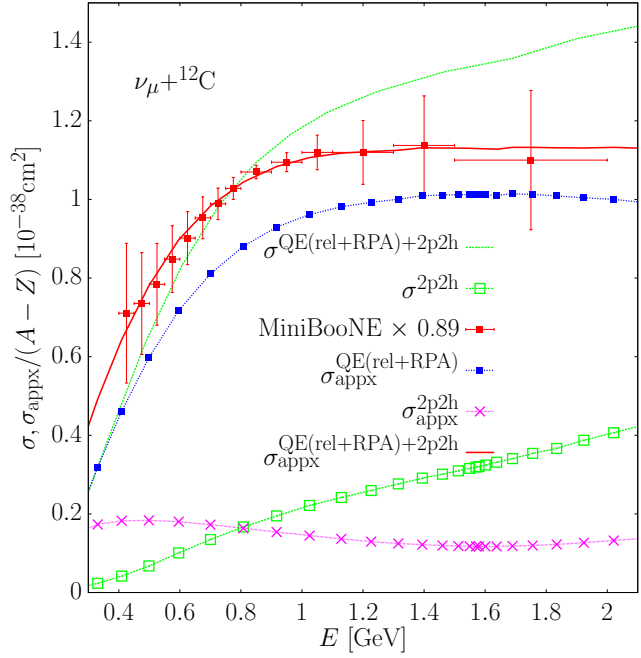}
\caption{Theoretical $\sigma$ and approximate $\sigma_{\rm appx}$,
 ( defined in Eqs.~(\ref{eq:poormodel2})--(\ref{eq:poormodel2b})) CCQE-like integrated cross sections in carbon as a
function of the neutrino energy. For
consistency with our previous results for the flux-folded double
differential  cross section $d\sigma/dE_\mu  d\cos\theta_\mu$ in Ref.~\cite{Nieves:2011yp}, 
the MiniBooNE data~\cite{AguilarArevalo:2010zc} and errors have been re-scaled by
a factor 0.89. The shape errors are shown for the MiniBooNE data.}
\label{fig:sigappx2}
\end{figure}

Taking into account that 
$d\sigma^{\rm 2p2h}/d\ereco(E;\ereco)$  is almost negligible
for $E\le \ereco$ and that it shows a quite long tail  above
this energy (see bottom
panel of  Fig.~\ref{fig:dsigdereco}), is then easy to understand the
redistribution of strength  from high  to low neutrino energies observed
in $\sigma_{\rm appx}^{\rm 2p2h}(E)$ when it is compared to the actual
$\sigma^{\rm 2p2h}(E)$ cross section. Up to some approximation the
area is conserved, though.

Finally, in  Fig.~\ref{fig:sigappx2} we compare the
MiniBooNE CCQE-like data with both $\sigma$ and $\sigma_{\rm
  appx}$. We see an excellent agreement between the latter one and the
data scaled down by a factor 0.89. As mentioned, our
QE(rel+RPA)+2p2h model successfully describes the MiniBooNE CCQE-like flux averaged double differential
cross section $d\sigma/dE_\mu d\cos\theta_\mu$ data, up to a global
scale $\lambda$ ($=0.89\pm 0.01$)~\cite{Nieves:2011yp}. This is the best observable
to compare with theoretical models because both the muon angle and energy are directly measured
quantities, and thus the shape of this distribution is readily
obtained from the number of events measured for each muon kinematical
bin. To obtain the absolute normalization of the distribution,
however, it is necessary to rely on some estimate for the number of incident neutrinos per unit
of area ($N_{inc}$). We believe the obtained value for
$\lambda \sim 0.89$ in \cite{Nieves:2011yp} indicates that the actual
number of incident neutrinos per unit of area might be larger than the
central value assumed in the MiniBooNE analysis. This would be still
consistent with the MiniBooNE
estimate of a total normalization error of
10.7\%~\cite{AguilarArevalo:2010zc}. The value of $N_{inc}$ is
needed to estimate $\langle \sigma \rangle_{\rm Exp}$ in
the expression of Eq.~(\ref{eq:poormodel}) for $\sigma_{\rm appx}(E)$,
and thus the predictions of our model in  Eq.~(\ref{eq:thmodel}) would
have to be multiplied by $1/\lambda$, or equivalently the data have to be 
scaled down by a factor $\lambda$.

Coming back to the results displayed in Fig.~\ref{fig:sigappx2},  
we should conclude that the unfolded cross section published in
\cite{AguilarArevalo:2010zc} appreciably differs from the real
one $\sigma(E)$. Actually, it is not a very clean observable after noticing the
importance of multinucleon mechanisms, because the unfolding itself is
model dependent and assumes that the events are purely QE. The same
limitation occurs for the differential cross section $d\sigma /dq^2$,
given that $q^2$ is also deduced assuming the events are QE. When
compared with the ``real`` $\sigma(E)$, the MiniBooNE unfolded cross
section exhibits an excess (deficit) of low (high) energy neutrinos,
which is an artifact of the unfolding process that ignores multinucleon
mechanisms.

The semi-phenomenological model of
Refs.~\cite{Martini:2009uj,Martini:2010ex} predicts a theoretical
cross section $\sigma(E)$ that provides a  good
description of the MiniBooNE unfolded cross section of
Ref.~\cite{AguilarArevalo:2010zc}. However, we have shown here that
these data do not correspond to the actual cross section because the
unfolding process is biased. To compare with these
data, the authors of \cite{Martini:2009uj,Martini:2010ex}  should
carried out a procedure similar to that proposed in
Eq.~(\ref{eq:poormodel}). Since the model of these works
includes strong 2p2h contributions, we would expect an  appreciable change in the
shape, as discussed above, that might distort quantitatively and
qualitatively the agreement with the MiniBooNE unfolded cross section 
found in Refs.~\cite{Martini:2009uj,Martini:2010ex}. 

In Ref.~\cite{Martini:2012fa}, Martini et al. have paid special
attention to the flux dependent $P(E| \ereco )$ probability. We
however believe that $P(\ereco| E)$ (or equivalently the differential
cross section $d\sigma/d\ereco$) could be more illuminating.  First,
because it does not depend on the neutrino flux, and second because it
can be used by experiments to determine $\sigma(E)$ thanks to Bayes's
theorem.  Finally, we should also mention a recent and quite
comprehensive work on neutrino-nucleus observables~\cite{Mosel:2012hr}
has also found, using some simple models for multinucleon mechanisms,
that 2p2h interactions lead to a downward shift of the reconstructed
energy in agreement with our results.

\section{Conclusions}

We have shown that because of the the multinucleon mechanism effects,
the algorithm used to reconstruct the neutrino energy is not adequate
when dealing with quasielastic-like events. This effect is relevant to neutrino oscillation 
experiments that uses the CCQE samples to compute the neutrino energy. 
The assumption of a pure CCQE interaction introduces biases in the 
determination of $\Delta m^2$ and mixing angle.  Moreover, 2p2h
contributions put also limitations on the validity of the flux
unfolding procedure used in \cite{AguilarArevalo:2010zc}. 
The MiniBooNE unfolded cross section exhibits an excess (deficit) of
low (high) energy neutrinos, which is an artifact of the unfolding
process that ignores multinucleon mechanisms. Actually,
$\sigma_{\rm appx}$, defined in
Eqs.~(\ref{eq:poormodel2})--(\ref{eq:poormodel2b}), provides an
excellent description of the data of
\cite{AguilarArevalo:2010zc}. This, together with our previous results
in Ref.~\cite{Nieves:2011yp} for the CCQE-like flux averaged double differential
cross section $d\sigma/dE_\mu d\cos\theta_\mu$, make us quite
confident on the reliability of our QE(rel+RPA)+2p2h microscopical model derived in
Refs.~\cite{Nieves:2004wx,Nieves:2011pp}. Furthermore, because it is just a
natural extension  of previous successful studies of photon,
electron, and pion interactions with nuclei~\cite{Gil:1997bm,Carrasco:1989vq,Nieves:1991ye,Nieves:1993ev}.

\appendix
\section{Relativistic vs non-relativistic QE total CC neutrino cross
  sections within the scheme of Ref.~\cite{Nieves:2004wx}}
  
\label{sec:app}
 The  left panel of Fig. 18 in Ref.~\cite{Nieves:2011pp} corresponds to
 the flux-unfolded MiniBooNE $\nu_\mu$ CCQE-like cross section per
 neutron as a function of the neutrino energy. There, both the QE
 and the 2p2h contributions to the total cross section were also shown
 separately. The latter ones were 
 computed fully relativistically, while  the QE predictions were taken 
from a previous work~\cite{Nieves:2004wx}. Concretely, results 
that included RPA and FSI effects, were selected and displayed in  the
Fig. 18. Since the approach
used in \cite{Nieves:2004wx} to account for FSI effects was not
relativistic, the QE curves displayed in (both panels) Fig. 18 of
Ref.~\cite{Nieves:2011pp} neglect some relativistic effects for the
nucleons.  In particular, those results are based on a non-relativistic
approximation for the nucleon (particle and hole)
propagators\footnote{Actually, all QE results of
Ref.~\cite{Nieves:2011pp} were obtained with non-relativistic nucleon
propagators.}. This is one of the sources of systematic errors, among
others, that should be accounted for by the bands of theoretical
uncertainties displayed in Fig. 18 (see the discussion of the third
paragraph of pag. 16 in \cite{Nieves:2011pp}). The use of
non-relativistic nucleon propagators is responsible of the odd
behaviour of the QE results in the Fig.18 of 
Ref.~\cite{Nieves:2011pp} when the neutrino energy
increases. Indeed,  the QE cross sections 
are too big for neutrino energies above 0.5-0.6 GeV and these
predictions clearly depart~\cite{caballero} from the common pattern exhibited by
different models collected in a review talk presented in the NUINT
2009 Workshop~\cite{Boyd:2009zz}.  

\begin{figure}
\makebox[0pt]{\includegraphics[width=0.55\textwidth]{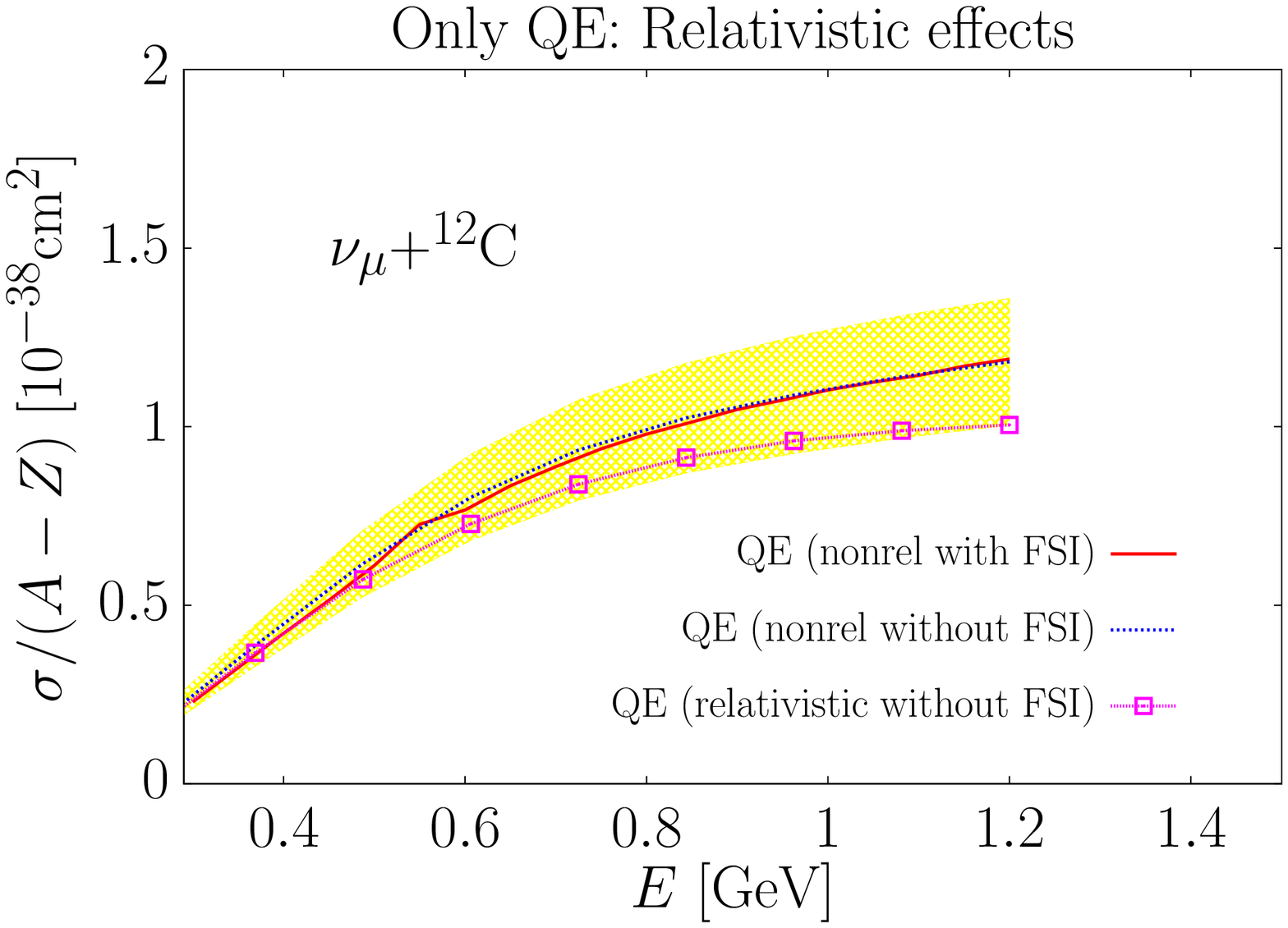}}\\ 
\vspace{0.5cm}
\makebox[0pt]{\includegraphics[width=0.55\textwidth]{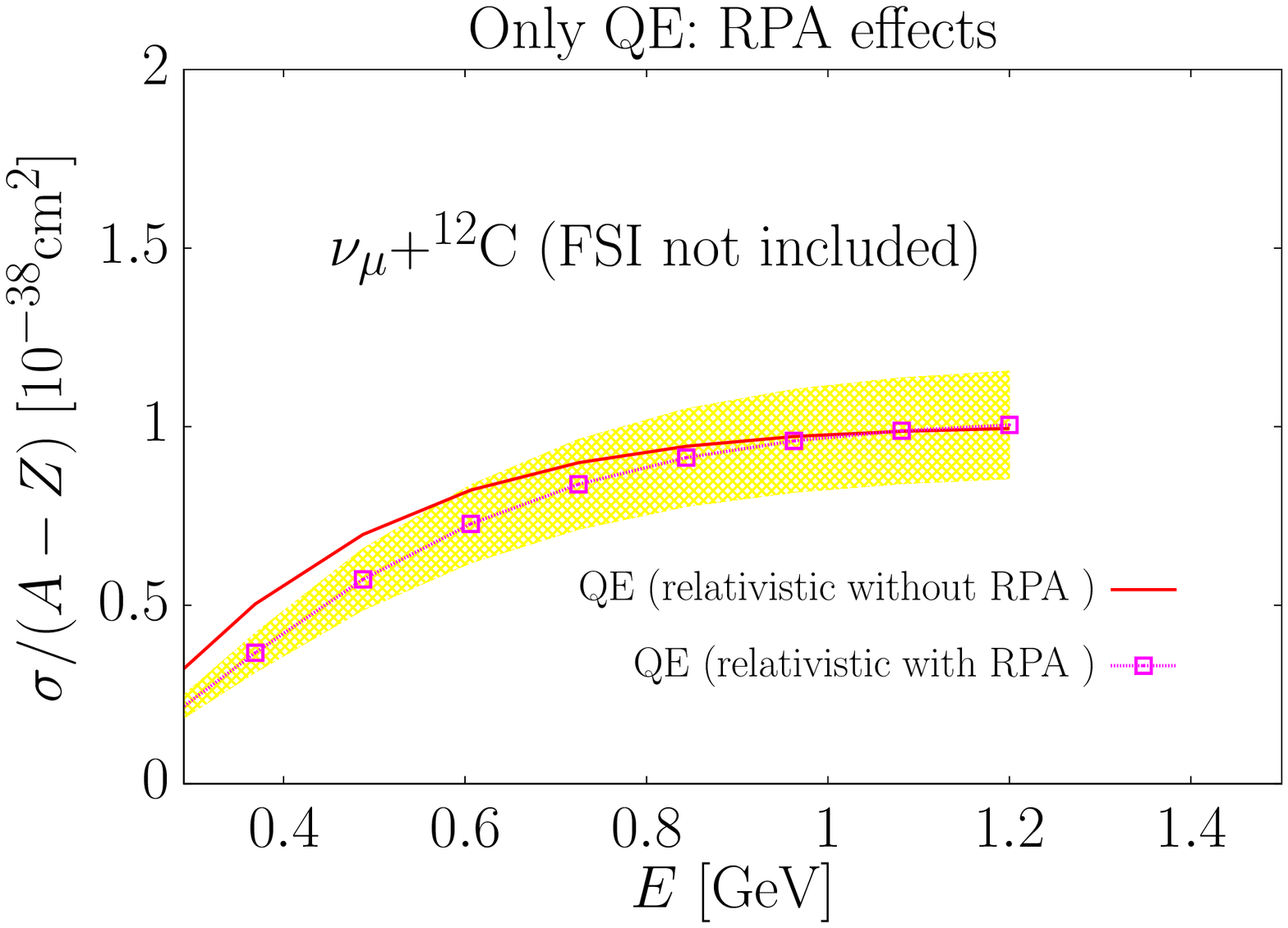}}
\caption{Different theoretical predictions for neutrino CCQE total
  cross section off $^{12}$C obtained from the model of
  Ref.~\cite{Nieves:2004wx}. Yellow bands account for a 15\% theoretical
  uncertainties that might affect the nuclear corrections included in
  the model of  Ref.~\cite{Nieves:2004wx}, as discussed in
  \cite{Valverde:2006zn}.}
\label{fig:fig1}
\end{figure}

In the top panel of Fig.~\ref{fig:fig1}, we show the size of FSI and
relativistic effects within the QE model of Ref. \cite{Nieves:2004wx}
(note that there, results obtained using relativistic nucleon
propagators, but neglecting FSI, were also shown). In all curves of
this top panel, RPA effects are taken into account. We see that FSI
have little effect on the integrated cross sections, though FSI might
affect differential distributions~\cite{Nieves:2004wx}, and we should
also pointed out that some relativistic approaches to FSI lead to
larger increases of the total cross section~\cite{Meucci:2011vd}. On
the other hand, for neutrino energies above 1 GeV, relativistic
effects for the nucleons reduce the cross sections by around 15\%, but
still the relativistic QE results lie within the theoretical error
band assumed for the QE predictions in Ref.~\cite{Nieves:2011pp}.

The flux folded CC double differential neutrino cross section,
measured by the MiniBooNE Collaboration, was analyzed in
Ref.~\cite{Nieves:2011yp} by using the full relativistic model of
Ref.~\cite{Nieves:2004wx} without the inclusion of FSI, but
taking into account RPA correlations, which effects in the
integrated flux-unfolded cross section can be seen in the bottom panel
of Fig.~\ref{fig:fig1}. This is the model that is used in this
work. Note, as can be appreciated in this latter plot, that though RPA
effects are quite relevant for low neutrino energies, they are
negligible above 1 GeV, and much smaller than the theoretical
uncertainties, discussed in \cite{Valverde:2006zn}, for $E > 0.7$
GeV.
\begin{acknowledgments}
 We thank R. Tayloe and G. Zeller for useful comments regarding the
data unfolding used by the MiniBooNE collaboration. This research was supported by DGI and FEDER funds, under contracts
 FIS2011-28853-C02-01, FIS2011-28853-C02-02, FIS2011-24149,
 FPA2011-29823-C02-02 and the Spanish Consolider-Ingenio 2010
 Programme CPAN (CSD2007-00042), by Generalitat Valenciana under
 contract PROMETEO/2009/0090, by Junta de Andaluc\'\i a grant FQM-225
 and by the EU HadronPhysics2 project, grant agreement no. 227431.
\end{acknowledgments}


\end{document}